\documentclass[sigconf]{acmart}
\renewcommand\footnotetextcopyrightpermission[1]{} 
\usepackage{tikz}
\usepackage{amsmath}
\usepackage{tabularx}
\newcolumntype{Y}{>{\centering\arraybackslash}X}

\newcolumntype{Z}{>{\hsize=1.2\hsize}X}
\newcolumntype{Q}{>{\hsize=.8\hsize}X}
\newcolumntype{V}{>{\hsize=.15\hsize}X}

\usepackage{caption}
\usepackage{subcaption}
\usepackage{stfloats}
\usepackage{multirow}
\usepackage{multicol}
\usepackage{makecell}
\usepackage{calc}  
\usepackage{enumitem} 
\usepackage{nicematrix,tikz}
\usepackage{xurl}

\usepackage{filecontents}

\AtBeginDocument{%
  \providecommand\BibTeX{{%
    \normalfont B\kern-0.5em{\scshape i\kern-0.25em b}\kern-0.8em\TeX}}}

\begin{document}

\title{Meaningful Context, a Red Flag, or Both? Users' Preferences for Enhanced Misinformation Warnings on Twitter}

\author{Filipo Sharevski}
\affiliation{%
  \institution{DePaul University}
  \streetaddress{243 S Wabash Ave}
  \city{Chicago}
  \state{IL}
  \postcode{60604}
    \country{United States}
}
\email{fsharevs@cdm.depaul.edu}

\author{Amy Devine}
\affiliation{%
  \institution{DePaul University}
  \streetaddress{243 S Wabash Ave}
  \city{Chicago}
  \state{IL}
  \postcode{60604}
    \country{United States}
}
\email{adevine@depaul.edu}

\author{Emma Pieroni}
\affiliation{%
  \institution{DePaul University}
  \streetaddress{243 S Wabash Ave}
  \city{Chicago}
  \state{IL}
  \postcode{60604}
    \country{United States}
}
\email{epieroni@depaul.edu}

\author{Peter Jachim}
\affiliation{%
  \institution{DePaul University}
  \streetaddress{243 S Wabash Ave}
  \city{Chicago}
  \state{IL}
  \postcode{60604}
    \country{United States}
}
\email{pjachim@depaul.edu}

\renewcommand{\shortauthors}{Authors}

\begin{abstract}
Warning users about misinformation on social media is not a simple usability task. Soft moderation has to balance between debunking falsehoods and avoiding moderation bias while preserving the social media consumption flow. Platforms thus employ minimally distinguishable warning tags with generic text under a suspected misinformation content. This approach resulted in an unfavorable outcome where the warnings ``backfired'' and users believed the misinformation more, not less. In response, we developed enhancements to the misinformation warnings where users are advised on the context of the information hazard and exposed to standard warning iconography. We ran an A/B evaluation with the Twitter's original warning tags in a 337 participant usability study. The majority of the participants preferred the enhancements as a nudge toward recognizing and avoiding misinformation. The enhanced warning tags were most favored by the politically left-leaning and to a lesser degree moderate participants, but they also appealed to roughly a third of the right-leaning participants. The education level was the only demographic factor shaping participants' preferences. We use our findings to propose user-tailored improvements in the soft moderation of misinformation on social media.
\end{abstract}




\maketitle

\section{Introduction}
Warnings and secure user behavior seems to have a perennially fraught relationship, despite the rich mediation of usability \cite{Egelman, Akhawe}, interaction/visual design \cite{Felt}, and behavioral insights \cite{Wash, Roberts}. It is understandable that the complexity of this problem requires patience and eventual alignment between the security literacy of the average user and the pace with which new security hazards are introduced into users' daily life \cite{Fagan, Herley}. Usable security has, for one, made noticeable advancements of warnings that users do actually heed in conformance with the security recommendations: avoiding phishy websites and questionable attachments \cite{Reinheimer}, skipping unencrypted communication \cite{Tiefenau}, warming up to multi-factor authentication \cite{Krombholz}, and following up on system updates \cite{Mathur}. Advancements such as adaptive strategies for getting accustomed to warnings and security advice also help users transition to an acceptable secure behavior \cite{Forget, Hielscher}. 

What actually is a bit difficult to understand is why, despite these advancements in usable security, warnings about misinformation on social media have made little progress in fostering desirable security behavior \cite{cose2021}. One could argue that the nature of the security hazard differs between the two settings - traditional programmatic security is far more complex to grasp than picking up on a causal post that links the COVID-19 vaccines with infertility - and that makes designing misinformation warnings an entirely different challenge. True, the one-size-fits-all here won't work because yesterday were the elections \cite{Jachim, Zannettou}, today is COVID-19 and QAnon \cite{Pieroni, Ahmed}, and who knows what alternative narratives will emerge tomorrow. Embracing this predicament as a challenge in a usable security context has been sporadic so far, with the focus largely placed on mapping the ``sources of misinformation'' \cite{Dupuis}.

Misinformation sources won't go away. They existed long ago, learned to adapt and thrive in new information environments, and so long as the Internet evolves, they will too \cite{Rid}. In the context of social media, these sources generate \textit{misinformation} content that includes all false or inaccurate information such as: disinformation, fake news, rumors, conspiracy theories, hoaxes, trolling, urban legends, and spam \cite{Wu}. It took some time for mainstream platforms to acknowledge that they have a serious problem on hands when misinformation started piling up \cite{Mosseri}. They responded with warnings which conformed to the aesthetics of their interfaces and with language presumably appearing as unbiased and non-judgmental to users with diverse perspectives \cite{Smith}. But this so-called ``soft moderation'' was applied halfheartedly, turning the warnings into \textit{hazards themselves}---users started believing the misinformation more, not less, when a warning was explicitly appended to it \cite{Clayton, Pennycook}. 

The design of the warnings, thus, requires adaptation of the approach to retain their usability in various misinformation scenarios while avoiding a ``backfire effect''  \cite{nspw2021}. To help this effort, we developed enhancements to the misinformation warning tags used by Twitter and evaluated them with a sample of 337 regular users. These enhancements address two elements that mainly contribute to the aforementioned predicament of soft moderation: meaningful context of the intentionally spread misinformation \cite{Wu} and sufficiently potent interruption of the regular social media consumption flow \cite{Gawronski, Ecker}. Therefore, we formulated the warnings' text to fit the scenario surrounding a misinformation tweet and introduced red flag watermarks as a characteristic iconography of the visual frictions that users encounter in every aspect of their daily life \cite{Cox}. 

The results of an A/B evaluation study with the warning tags currently employed on Twitter show that the majority of users \textit{do} welcome the usable security enhancements. The added meaningful context was praised in helping participants avoid, ignore, and skip  misinformation \textit{``right away.''} The red flag watermarks were lauded for their \textit{``attention-grabbing''} effect. Expectantly, there were also groups of users that leveraged this opportunity to express their protest against soft moderation as a way of forceful opposition-opinion-forming by Twitter as a self-appointed truth authority. Therefore, we analyzed the sentiment the warning tags incited and found that the enhancements did tilt the overall sentiment toward more positive from the \textit{status quo} of the original Twitter warnings. 

Sentiment often reflects users' political leanings and is shaped by the structure of their demographic identity \cite{Thorson, Reuter}. Our results suggest that the left-leaning participants overwhelmingly welcomed the meaningful context, and the moderates and right-leaning joined them in lauding both the context and the red flag watermarks. Users' age, gender, and race/ethnicity did not factor in the sentiment in a significant measure; only the education level did. While users with either a high school education/GED or a college diploma were evenly split in preferring the original warnings and their enhanced counterparts, the users with some college education overwhelmingly preferred the latter. All but one of the users with a post-graduate level of education were entirely in preference for both the context and the red flag watermarks. 

\vspace{1em}

\noindent \textbf{Scope and contribution of this work.} With this work we aim to materialize the wealth of usable security cues, nudges, and advises in a social media environment where  towards curbing misinformation. Our contributions, respectively, are:

\begin{description}
\item The first A/B evaluation of enhanced social media warnings providing meaningful context and introducing visual design frictions in interacting with misinformation;

\item Analysis of users' sentiment toward soft moderation in general and enhanced warning tags in particular from a political and demographic perspective;  

\item Basis and recommendations for user-tailored adaptations of soft moderation toward mindful and safe interaction on social media.
\end{description}

Following this introduction, we delve into the current state of misinformation warnings on social media in Section \ref{section:2}. We then elaborate on our usable security enhancement approach in Section \ref{section:3}. Section \ref{section:4} provides the results of our A/B evaluation study and sentiment analysis. We discuss the results in Section \ref{section:5} and provide our recommendations for the future of soft moderation before we conclude the paper in Section \ref{section:6}.

\section{Achtung! Misinformation} \label{section:2}
Warnings on social media usually come in two main forms: (i) \textit{interstitial covers} which obscure the misinformation and require users to click through to see the information; or (ii) \textit{trustworthiness tags} which appear under the content and do not interrupt the user or compel action \cite{Kaiser}. The former are more suitable for sensitive content where the exposure to the hazards should be avoided in the first place  and the latter are usually applied to disputed or unverified content where the decision whether it is or not of misinformation nature is left to the user. But the COVID-19 ``infodemic'' demanded all hands on deck for soft moderation, and so mainstream social media platforms applied both warning variants to warn users of misleading and harmful COVID-19 information \cite{Roth, Mosseri}. 

Evidence suggest that only the interstitial covers, but not the trustworthiness tags, make the users heed the warnings of misinformation \cite{cose2021, Sanderson, Zannettou, Saltz}. It is tempting to simply discard the trustworthiness tags and only use interstitial covers, however. The interstitial covers do require additional clicks to get to the content in question, which could make the users avoid the content, but leave with a feeling that the social media platform is overtly imposing, ``biased,'' ``punitive'' or ``restrictive of free speech'' \cite{Saltz, Jhaver}. The trustworthiness tags might be more usable and mitigate the overt intrusion by blending with the visual aesthetic of the platforms interface (e.g. same colors, fonts, and obscure text), but they do run into other problems. Next to the ``backfiring effect'' \cite{Clayton, Nyhan, Ecker, Thorson}, the tags could desensitize users to soft moderation when applied too frequently or create an ``illusory truth effect'' \cite{Pennycook1}. The absence of the tags in some scenarios might even create an `implied truth effect'' and lead users deem any misinformation content they encounter as credible and accurate \cite{Pennycook}.

Other factors also contribute to these negative effects, for example users' political affiliations and demographic identities. When trustworthiness tags directly challenged political falsehoods, they had the intended effect on Democrats but the opposite effect (e.g. they ``backfired) on Republicans \cite{Thorson}. In the context of the COVID-19 pandemic, the tags resulted in a ``belief echo,'' manifested as skepticism of adequate COVID-19 immunization particularly among Republicans and Independents \cite{cose2021}. Usually older users with a level of education corresponding to higher analytical thinking succumb less to these negative effects \cite{Reuter}. Another factor is the asymmetrical nature of soft moderation---the mere exposure to misinformation often generates a strong and automatic affective response, but the warning itself may not generate a response of an equal and opposite magnitude \cite{Gawronski}. This is because the trustworthiness tags often lack meaning, have ambiguous wording, or ask users to find context themselves which is cognitively demanding and time consuming \cite{Ecker}. Therefore, a natural step toward minimizing the said negative effects, is enhancing the trustworthiness tags to counter this asymmetry while keeping the appeal relevant for users of all ages, analytical prowess, and political leanings \cite{Reuter}.

\section{Enhanced Misinformation Warnings} \label{section:3}
The trustworthiness tags applied by Twitter make an interesting case of usable security interventions. Appended under suspected misinformation, this brand of tags warns \textit{after} a user is exposed to the potentially harmful content \cite{Roth}. Choosing to warn a user after-the-fact goes somewhat against the practice of using warning screens in browsers that come \textit{before} a user gets a chance to visit a questionable website \cite{Felt} (this effect is achieved with the interstitial covers, but they are verbose and disruptive of the natural social media consumption flow \cite{Bandy}). One could argue that the after-the-fact notification is chosen to counter ``habituation'', or the diminished response with repetitions of the same warning screens like these, or perhaps break the effect of ``generalization'' that might occur when habituation to these screens carries over to novel security interventions that look like the warning tags \cite{Vance}. Seemingly designed to camouflage itself amongst the existing interface features, the warning tags are \textit{blue} and not \textit{red} in color, they do not obscure the suspected misinformation tweet, nor do they occur predictably like the warning screens every time an Internet browser cannot verify the visiting website's certificate (the tweets in question have to be fact checked, if not automatically flagged \cite{Jachim}).  

Twitter's warning tags might compare to the lock icons at the beginning of an URL bar in a browser indicating a ``secure'' connection \cite{Reeder}. Besides the habituation and generalization, the lock icons are confusing and don't convey the threat to the users in the first place so proposals have been made to pair the usable security iconography with words when possible \cite{Porter-Felt}. Thus, it seems reasonable to pair an exclamation mark with a generic short text for warning users about misinformation tweets. But both the icon and the text are colored in the specific Twitter blue and fail to provide contrast to attract user's attention like the lock icons do with either red for ``insecure'' or green for ``secure'' browsing (alternatively a display of a locked/broken golden lock or strike-through the word ``HTTPS''). Deliberately avoiding contrast makes it easier for users to overlook, ignore, or simply mistrust the warning tags as honest security aids \cite{Krombholz}.

Perhaps pairing the generic warning text with a link to a Twitter-curated page or external trusted source containing additional information on the claims made within a suspected tweet could compensate for the lack of contrast. Often with a one-liner, users are offered for example to ``get the facts on the COVID-19 vaccine,'' ``learn why health officials say vaccines are safe for most people,'' or ``learn how the voting by mail is safe and secure'' \cite{Roth}. The Fear of Missing Out (FOMO) aside \cite{Alutaybi}, the warning text in fact \textit{advises} users to contextualize the tweet themselves on the particular (mis)information topic. Users, unfortunately, rarely heed this advice and largely refuse to investigate any (mis)information further \cite{Geeng}. 

Security advice is not entirely anathema to users, particularly when it comes to their online security hygiene \cite{Redmiles}. So it is not unreasonable to expect that users might heed the suggestion brought forth, brandished in a warning tag, if the advice itself provides a \textit{meaningful context} for a particular topic of contention on Twitter without asking users to follow a link (which conflicts Twitter's own idea of curating ``more accounts, and less links'' in user's feeds \cite{Bandy}). Balancing for comprehensibility, we developed enhanced warning tags that provide meaningful context in regards (1) fabricated facts; and (2) improbable interpretations of facts. The enhancement choice follows the misinformation front put forth by Twitter and allowed us to conduct an A/B usability evaluation with the current warning tags applied to misinformation hazard. The enhanced warnings, in their tag-only variant, incorporate catchy acronyms as frictions indented to grab users attention in the absence of contrast ~\cite{Cox}. 

We paired the text-only warning tags with the hereto ignored usable security intervention when it comes to misinformation: red flags as watermarks over suspected misinformation tweets. The tag-and-watermark variant provided option for us to also test users' receptivity to warnings that incorporate contrast (red), gestalt iconography for general warnings (flag), and actionable advice for inspection (watermark). The choice of red flag was made after an extensive deliberation concerning warning design \cite{Wogalter, Leonard}, warning cognition (automatic or System I; deliberate or System II) \cite{Moravec}, and user experience design \cite{Felt}. We decided against a smaller red flag as smaller labeling symbols were ignored on social media, e.g. Facebook used a small red box on the left with an exclamation mark and was either ignored or users believe the flagged post more, not less \cite{Roozenbeek}. We decided to use red and not other color flags because a ``red flag'' is a common signal of oncoming danger and requires users to switch from System I to System II of cognition. Green usually signals ``no danger'' while orange or yellow signal ``caution'' but are often processed by System I cognition \cite{Rogers}. Red also has the highest ``perceived hazardousness'' on the color palette ~\cite{Wogalter}.


\begin{figure*}[!t]
\subfloat[Text-only SPAM Warning Tag \label{fig:SPAM-A}]{%
  \includegraphics[width=.45\linewidth]{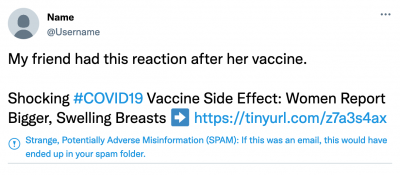}%
}\hfill
\subfloat[Text-and-flag SPAM Warning Tag\label{fig:SPAM-B}]{%
  \includegraphics[width=.45\linewidth]{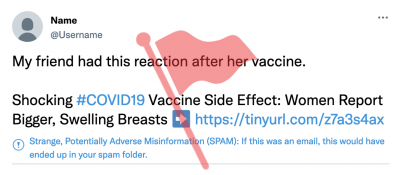}%
}
\caption{Warning Tags Contextualizing Fabricated Facts}
  \label{fig:SPAM}
\end{figure*}

\subsection{Fabricated Facts (SPAM)}
The first text-only warning tag is shown in Figure \ref{fig:SPAM-A}. We crafted a tweet, based on \cite{tweet_corvus}, and tagged for fabricated facts and presented it under a generic name, username, and without a profile image to avoid any threat to the validity of our A/B evaluation. Instead of advising the users to ``get the facts about the COVID-19 vaccine''\cite{Roth}, we coined a catchy, yet familiar acronym: \textbf{SPAM} or \textbf{S}trange, \textbf{P}otentially \textbf{A}dverse \textbf{M}isinformation. With \textbf{SPAM} we wanted to see if we can contextualize the tweet's content, with an analogy to an already meaningful aspect of spam email, something most Twitter users have experience with \cite{Brunton}.

We did break the one-liner rule for the warning text, but we opted for a minor engagement pain for a major gain in increased attention and warning adherence behavior. Our warning text following the \textbf{SPAM} acronym read: ``\textbf{If this was an email, this would have ended up in your spam folder}.'' The overarching idea  with the \textbf{SPAM} warning was to harness the ``availability'' and ``recognition'' heuristics characteristic captured in a Twitter flow \cite{Nielsen, Acquisti}. Misinformation and fabricated facts are not always spam or vice versa, but anyhow align on the actionable outcome: ignore, delete, or take it with a grain of salt \cite{Redmiles18}, which we argue is preferable compared to the ``backfiring effect'' of the generic warning tags \cite{Clayton, cose2021}.

The upgraded \textbf{SPAM} warning tag with a 50\% transparency red flag watermark over the entire tweet is shown in Figure \ref{fig:SPAM-B}. The ``upgrade'' bolsters the warning tag context along the same lines of ``availability'' and ``recognition'' heuristics by invoking the well-known analogy between red flags and calls for attention. We opted for a watermark and not a replacement of the exclamation point inline the warning tag to avoid confusion with the red flag emoji frequently used on social media. The watermarking, centered in a ratio over the entire tweet area, follows the paradigm for misinformation flagging proposed in \cite{nspw2021} with a midpoint transparency to create a non-negligible design friction for anyone attempting to read the tweet. By this choice, we wanted to stretch the overall text-and-flag warning \textit{throughout} the suspected misinformation tweet and not only \textit{after} it.

\subsection{Improbable Interpretations of Facts \textbf{(FFS)}}
\begin{figure*}[b]
\subfloat[Text-only FFS Warning Tag \label{fig:FFS-A}]{%
  \includegraphics[width=.45\linewidth]{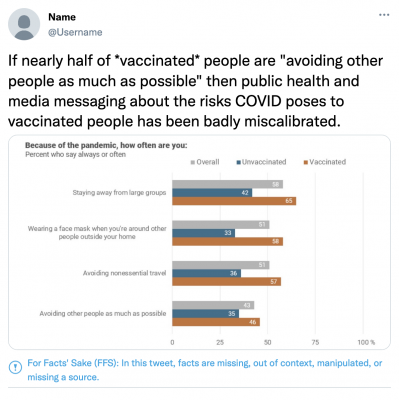}%
}\hfill
\subfloat[Text-and-flag FFS Warning Tag\label{fig:FFS-B}]{%
  \includegraphics[width=.45\linewidth]{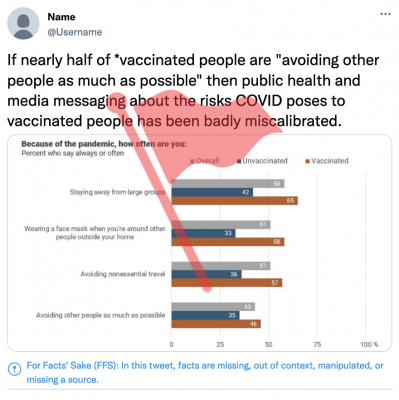}%
}
\caption{Warning Tags Contextualizing Improbable Interpretation of Facts}
  \label{fig:FFS}
\end{figure*}

The second set of warning tags is shown in Figure \ref{fig:FFS-A} and Figure \ref{fig:FFS-B} for the text-only and text-and-flag variants, respectively. Here we crafted a tweet, based on \cite{tweet_nate_silver}, containing an improbable interpretation of facts, keeping the engagement and posting structure in the similar order. In this case, we chose to provide a meaningful choice of context when tweets attempt to ``spin'' facts as a refined way of promulgating misinformation \cite{Fallis}. This practice, for example, earned Representative Marjorie Taylor Greene a permanent ban from Twitter \cite{Alba}. Since we want to draw users' attention to such practices, we decided to ask whether they consider such tweets for \textbf{F}or \textbf{F}acts' \textbf{S}ake or \textbf{FFS}, if not for anything else. We deliberately selected the acronym \textbf{FFS} to blend with the characteristic communication on Twitter that utilizes ``compact language'' due to the tweets' length restriction \cite{Zimbra}. 

The \textbf{FFS} warning tag intended to provoke a pause in ``recognition'' heuristics since there are multiple meanings associated this acronym. We were aware that this might cause brief confusion, but nonetheless proceeded, since we wanted to explore if a brief confusion followed by contextual advice would suffice in refraining from taking the improbable interpretations of facts at face value. We utilized the growing evidence of ``design frictions''  purposefully created to disrupt mindless automatic interactions, prompting moments of reflection \cite{Cox}. The brief confusion, promptly, is resolved by the following warning text advising users that ``\textbf{In this tweet, facts are missing, out of context, manipulated, or missing a source.}'' To gauge the limits of the warnings-as-friction, the red flag watermark provides another stimulus to capitalize on by seeing what works as a resolution against the questionable content: incomplete factual presentation \cite{Walter}, lack of contextual consistency \cite{Shan}, overt factual manipulation \cite{Shahi}, or obscure factual provenance ~\cite{Hounsel}.

\section{Evaluation Study} \label{section:3}
\subsection{Research Questions}
The evaluation of the enhanced warning tags was intended to gauge a preferential approach to soft moderation as well as understand the underpinning reasoning for it's acceptance (or lack thereof). A/B testing is a regular practice in usable security studies that informs the design of interface affordances, cues, and frictions \cite{Ruoti, Lyastani, Seiler-Hwang}. Building on the exposure to contextual warning tags, a qualitative inquiry of how they fare in the misinformation front is important because the soft moderation employed by social media in general, and Twitter in particular, so far has yielded far from desirable results \cite{Kirchner}. Users' often materialize their identity and political personas within social media and Twitter discourse \cite{Golbeck, jimei2022}, therefore we also investigated how this materialization shapes the preferences for our proposed soft moderation nudges. Based on this argumentation, the resulting research questions were: 

\begin{itemize}
    \item \textbf{RQ1:} What are the preferences of Twitter users for the \textbf{SPAM} and the \textbf{FFS} enhanced misinformation warning tags in both the text-only and text-and-flag variants? 
    \item \textbf{RQ2:} How effective are the \textbf{SPAM} and the \textbf{FFS}, enhancements in dispelling fabricated facts and improbable interpretations of facts? 
    \item \textbf{RQ3:} What is the relationship between the Twitter users' preferences for the enhanced misinformation warning tags and users' political leanings? 
    \item \textbf{RQ4:} What is the relationship between the Twitter users' preferences for the enhanced misinformation warning tags and users' demographic identities (race/ethnicity, level of education, gender identity, age)? 
\end{itemize}

\subsection{Recruitment}
Our study was approved by our Institutional Review Board (IRB) before any research activities began. Subsequently we set to sample a population that was 18 years or above old, regular Twitter users from the United States through the Amazon Mechanical Turk. Both reputation and attention checks were included to prevent input from bots and poor responses. The survey took around 20 minutes and participants were compensated with the standard participation rate (\$18 per hour). Participants were randomly assigned to either the A/B evaluation of the text-only or text-and-flag enhanced warning variants. The survey was anonymous and allowed users to skip any question they were uncomfortable answering.

We refrained from exposing the participants to similar stimuli to prevent from generalization and obtain a direct comparison to the original warning tags on Twitter. We also randomized the order of each of the \textbf{SPAM} and \textbf{FFS} text-only and text-and-flag segments for each participant. We selected the content of the tweets to be of relevance to the participants so they could meaningfully engage with the tweet’s content and see a clear relationship between the tweet and the warning tag (i.e. to prevent arbitrary and irrelevant responses). The two COVID-19 related tweets represent the main target of soft moderation front by Twitter during the execution of the study [November 2021 - January 2022] \cite{tvm}. We selected one misleading tweet by Nate Silver \cite{tweet_nate_silver}, and wrote a second one based on a common piece of vaccine misinformation \cite{tweet_corvus}. To account for accessibility, we provided alternative text describing each of the tweets and interventions we used to avoid visual misinterpretation. We assumed participants understood the Twitter interface, the tweets, and the warning tags.

\subsection{Method and Instrumentation}
Participants first indicated the reasons they usually come to Twitter for. Next, each participant was asked to indicate if they encountered warning tags and what were the content and the warnings about. We were aware that not every participant might have been exposed to warning tags so we included a small training segment where we created exposure to the concept of soft moderation with generic warning tags. The pre-exposure training, shown in the Appendix, was used to ensure a baseline understanding of content moderation among the participants, i.e. that Twitter uses content indicators for various types of contents (misinformation, sensitive content, graphic content, etc.). The training was general and referred to only ``content indicators'' without any references to ``misinformation to avoid any potential impact on user responses.

Participants then were asked to evaluate each of the enhancements in comparison to the original tag (``Get the facts about COVID-19'') \cite{Roth}. Participants were next asked if seeing an enhanced warning tag would influence their dismissal of the tweet or tweets on the same contested topic as misinformation. Finally we collected participants' political leanings, race/ethnicity, level of education, gender identity, and age. 

The qualitative responses were coded and categorized in respect the preference and the justification for it. These categories later helped perform a chi-square statistical analysis $\chi(n)$ of the relationships between the preferences and participants' political leanings as well as their demographic identities. We performed a basic exploratory analysis of the preferences and justification to uncover the aspects in which the enhanced warning tags fair well (or vice versa) as a usable security nudges against misinformation. For each of the justifications in the open-ended questions we performed a sentiment analysis using the Valence Aware Dictionary for Sentiment Reasoning (VADER) \cite{VADER, Dutta, Joseph}. VADER yields a compound score between -1 for a very negative piece of text, and 1 for a very positive one. We also used a Linguistic Inquiry and Word Count (LIWC) analysis to qualify the sentiment expressions in the responses respective to \textit{clout} and \textit{tone} \cite{Pennebaker}. Each one ranges between 0 and 100 with scores close to 0 indicates less confidence and weak argumentation (clout) or negative emotions (tone). 



Finally we performed a Correspondence Analysis (CA) on a contingency table with rows of adjectives/verbs as keywords and the justification text as columns. The CA projects the variance in justification onto two dimensions using a weighted single value decomposition \cite{Greenacre}. In CA, the further away the keywords are from the origin of the plot, the more discriminating they are, and smaller angles between a pro/against preference and a keyword (connected through the origin) indicates an association of the two. In our case, the two axes correspond with justifications' keywords (y-axis) respective to the participants' pro/against preferences (x-axis). 




\section{Results} \label{section:4}
After the consolidation and consistency checks, a total of 337 participants have completed the study, with 176 in the text-only and 161 in the text-and-flag warning tag groups, respectively. Users indicated that communication was the most frequent factor for coming to Twitter (85.4\%), followed by entertainment/cultural awareness (71.8\%), news (63.5\%), politics (46.5\%) and health (26.7\%). Around every third participant (32.9\%) has encountered some form of a warning tag as part of Twitter's soft moderation effort in general.

The distribution of participants per their self-reported political leanings was: 147 (43.6\%) left-learning, 96 (28.5\%) moderate, 61(18.1\%) right-leaning, and 33 (9.8\%) apolitical. In respect to race and ethnicity, 247 (73.3\%) identified as White, 29 (8.6\%) as Black or African American, 42 (12.5\%) as Asian, 12 (3.6\%) as Latinx, 3 (0.9\%) as Native Hawaiian or Pacific Islander, and 4 (1.1\%) as Other. Education-wise, 71 (21.1\%) of the participants had a high-school level, 57 (16.9\%) some college, 175 (51.9\%) 2- or 4-year college, and 57 (10.1\%) had a gradate level of education. Gender-wise, 154 (45.7\%) of the participants were female, 169 (53.4\%) were male, and 3 (0.9\%) identified as non-binary. Age-wise, 9 (2.7\%) were in the 18 - 24, 87 (25.8\%) in the 25 - 34, 136 (40.3\%) in the 35 - 44, 64 (19\%) in the 45 - 54, 33 (9.8\%) in the 55 - 64, and 8 (2.4\%) in the 65 - 74 bracket. 

To ensure consistency in the analysis and validity of the results, each of the open-ended responses in the survey was coded independently by three researchers. The codebook was simple and included a coding on the preference expressed for the A/B evaluation as well as codes for the preference justification quotes from the participants. The Fleiss's kappa $\kappa$, as a measure of inter-coder agreement, was 0.960 on average with a 0.878 lower bound for the 95\% confidence, which indicates an ``excellent'' inter-coder agreement overall.

\subsection{Fabricated Facts (SPAM)}
\subsubsection{A/B Evaluation}
The breakdown of preferences for both variants of the \textbf{SPAM} warning tag is given in Table \ref{tab:SPAM-Preferences}. In the text-only variant, more than a half of the participants who preferred the original warning tags explicitly echoed a protest against Twitter's intrusion in contested matters such as COVID-19 vaccination. Verbosity and confusion was cited by roughly one out of five participants as a preference against the \textbf{SPAM}. The same number of participants didn't provided any justification. A small number of participants judged the \textbf{SPAM} tag as misaligned with Twitter's aesthetic and therefore, illegitimate. Neither of the text-only warning tags was the choice of 12.6\% of the participants. 

The \textbf{SPAM} text-only warning tag (Figure \ref{fig:SPAM-B}) received the highest preference (46.3\%). The meaningful context provided by the extended security advice was welcomed by 43.2\% of them indicating that \textit{``The SPAM explanation is a valid one, and makes sense in the context of the tweet's content.''} The on-point warning of questionable content was cited by 36.8\% in preference of  \textit{``a direct misinformation label right there without having to dig further into it.''} One tenth of the pro-\textbf{SPAM} participants found the acronym and the text \textit{catchy}, \textit{cheeky}, and positively attention-grabbing. Reluctance to follow the links in the original tag variant was cited by 6\% of the participants. Only 4\% didn't provide any justification. 

The pairing of the red flag with the \textbf{SPAM} warning tag was either too distracting or an indicator of Twitter's intrusion into the way content should be consumed. The preference against the text-and-flag \textbf{SPAM} tag was expressed in terms of \textit{``visual clutter that makes the tweet more difficult to read''}, \textit{``Doomsday level of importance''}, or \textit{``symbol of political hate''}. The pro text-and-flag \textbf{SPAM} tag participants welcomed the attention grabbing of the red flag suggesting that \textit{``the flag gets your attention; the text tells you it is misinformation - I tend skim when reading twitter posts and the other one is not as noticeable.''}. The enhanced context and the on-point warning for misinformation was preferred because \textit{`` the flag reinforces the positive information that the tweet is spam''}.

\begin{table*}[!h]
\renewcommand{\arraystretch}{1.0}
\small
\caption{\textbf{SPAM}: Preferences}
\label{tab:SPAM-Preferences}
\centering
\begin{tabularx}{\linewidth}{|c|c|l|X|}
\Xhline{3\arrayrulewidth}
\textbf{Option} & \textbf{Pct.} & \textbf{Justifications} & \textbf{Representative Quotes}\\\Xhline{3\arrayrulewidth}
\multicolumn{4}{|c|}{\textbf{Text-only Warning Tags}} \\\Xhline{3\arrayrulewidth}

\multirow{4}{*}{\shortstack[c]{\textbf{Original (A)} \\ (41.1\%)}} & 55.6\% & Twitter intrusion &  Telling me something is spam is an opinion concerning this topic and feels intrusive to trying to control my opinions. \\\cline{2-4}
& 19.4\% & Verbosity/confusion & Because it's simple and straight to the point. The SPAM is confusing and too wordy. \\\cline{2-4}
& 19.4\% & No justification & Warning Tag A. \\\cline{2-4}
& 5.6\% & Legitimate & Get the facts seems more legit to me. \\\Xhline{3\arrayrulewidth}

\multirow{4}{*}{\shortstack[c]{\textbf{SPAM (B)} \\ (46.3\%)}} & 43.2\% & Meaningful Context & Letting me know something is SPAM and dangerous is more useful than telling me where to find facts;  \\\cline{2-4}  

& 36.8\% & On-point warning & It tells me right away why Twitter marked it as misinformation so I don't have to wonder the reason on my own. I can also easily decide if I agree and move on or research further outside of Twitter. I love this and would be happy to see this on posts. \\\cline{2-4}


& 10\% & Attention Grabbing & ``B'' is better at catching the attention of the reader. ``A'' could just be a service announcement - it just isn't strong enough.\\\cline{2-4}

& 6\% & Link reluctance & It's more detailed and explains why its there without having to click on anything \\\cline{2-4}

& 4\% & No justification & Warning Tag B. \\\Xhline{3\arrayrulewidth}

\multicolumn{2}{|c|}{\textbf{Neither} (12.6\%)} & Neutral & I wouldn't utilize either.  \\\Xhline{3\arrayrulewidth}

 \multicolumn{4}{|c|}{\textbf{Text-and-Flag Warning Tags}} \\\Xhline{3\arrayrulewidth}
 \multirow{3}{*}{\shortstack[c]{\textbf{Original (A)} \\ (37.9\%)}} & 44.2\% & Distracting Flag & Seeing the red flag almost makes the tweet look like it is harmful or not true at all. It stands out too much. \\\cline{2-4}

 
 & 42.6\% & Twitter intrusion &  I would prefer the original warning tag. ``B'' is too opinionated and biased.\\\cline{2-4}
 & 11.6\% & No justification &  Warning Tag A \\\Xhline{3\arrayrulewidth}

\multirow{3}{*}{\shortstack[c]{\textbf{SPAM (B)} \\ (46.0\%)}} & 41.9\% & Attention Grabbing &  I prefer the red flag, as it is impossible to miss. I often read Twitter on my phone, while I take the dog out and such, so I find myself thinking I should look something up after reading a tweet, but then I get busy doing other things and don't follow up. \\\cline{2-4}


 & 25.7\% & Meaningful Context &  I like the red flag for sure - and the warning tag beneath gives a better commentary on why there was a red flag. \\\cline{2-4}

 & 22.9\% & On-point warning &  The red flag makes it very obvious that the material is potentially false and can't be trusted \\\cline{2-4}
 
 & 9.5\% & No justification &  Warning Tag B \\\Xhline{3\arrayrulewidth}

\multicolumn{2}{|c|}{\textbf{Neither} (16.1\%)} & Neutral & I wouldn't prefer any of them. \\\Xhline{3\arrayrulewidth}
\end{tabularx}
\end{table*}

\subsubsection{Sentiment Analysis}
\begin{figure}[htbp]
  \centering
  \includegraphics[width=\linewidth]{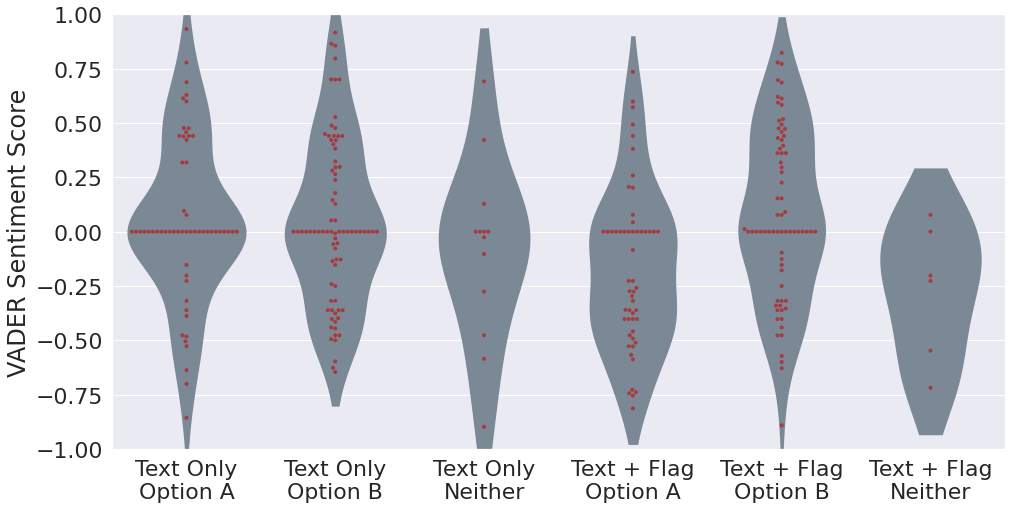}
  \caption{Sentiments: \textbf{SPAM} Warning Tags. The violin plots show a multimodal distribution of sentiments (number of responses) over the VADER sentiment score ranging from -1 (negative) to +1 (positive) sentiment.}
  \label{fig:q10_swarm}
\end{figure}

The sentiment analysis of the preferences for \textbf{SPAM} warning tags is shown in Figure \ref{fig:q10_swarm}. The violin plots show a multimodal distribution of sentiments where the original warning tag received an equal number of positive sentiments for being \textit{``simple and straightforward''} as well as negative sentiments that \textit{ ``rather not see Twitter's judgement on whether something is misinformation or not''.} The justifications showed low confidence (clout = 26.15) but positive emotions (tone = 60.65). The text-only \textbf{SPAM} positive sentiment outweighs the negative one that captures justifications indicating that \textit{```B' does a better job letting you know that the tweet's information is bad''}, with a bit more confidence (clout = 32.59) and on par with the positive emotions (tone = 62.74). 

The introduction of the red flag in the \textbf{SPAM} warning tag apparently induced more negative sentiment when justifying the choice for the original warning tag. The justifications were a bit more convincing (clout = 30.91) but the emotions were highly negative (tone = 7.61). The red flag increased the positive sentiment for the text-and-flag \textbf{SPAM} warning tag with the most confidence of all justifications (clout = 34.11) and positive emotions (tone = 55.52). While the participants that were neither ``A'' or ``B'' were evenly distributed in the text-only variant, the negative sentiment was dominant in the text-and-flag variant. Both being very low on confidence and high on negative emotions, the introduction of the flag might have exacerbated the feelings against the soft moderation for some of these participants.

Additionally, we performed a CA to review the adjectives used in explanations for user preferences for the \textbf{SPAM} warning tags. In the first component on the x-axis, which accounts for 56.57\% of the inertia in the justifications, all but three keywords show values larger than zero. This suggests a bit more consistency in the way that the preferences for both the ``A'' and ``B'' text-only options were worded. Put it simply, the predicative/comparative ``more,'' ``clear,'' and ``better'' adjectives were associated with the text-only \textbf{SPAM} tag, while the ``own'' and ``true'' with the original warning. The tendency for the prior is a praise of the enhancements themselves while the latter hints of a general contempt for soft moderation on Twitter. The less discriminating ``wrong,'' and ``false'' echo a similar sentiment by the neither ``A'' or ``B'' participants in the text-only variant.  The remaining keywords show values less than a zero, indicating that the adjectives used to justify those selections were generally less consistent, outside of the trend of expressing the preferences for text-and-flag \textbf{SPAM} with the keywords ``obvious'' and ``red.'' 


\begin{figure}[htbp]
  \centering
  \includegraphics[width=\linewidth]{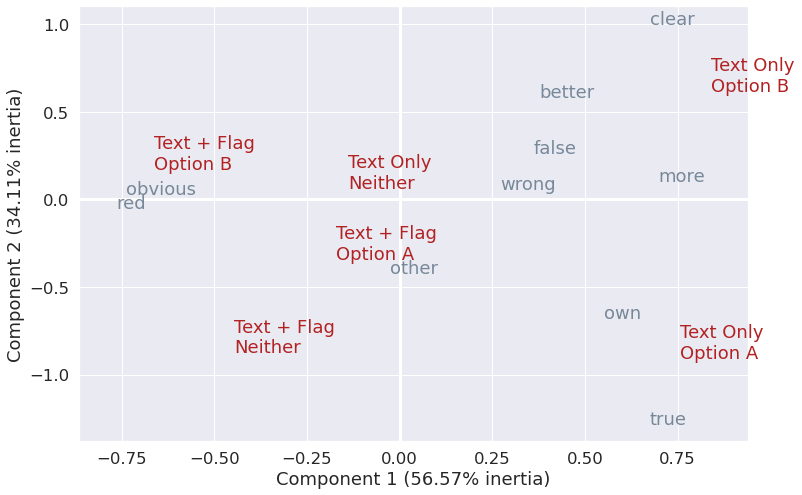}
  \caption{Correspondence Analysis: \textbf{SPAM} Warning Tags}
  \label{fig:q10_corr}
\end{figure}

\subsubsection{Dispelling Fabricated Facts}
\begin{figure}[htbp]
  \centering
  \includegraphics[width=\linewidth]{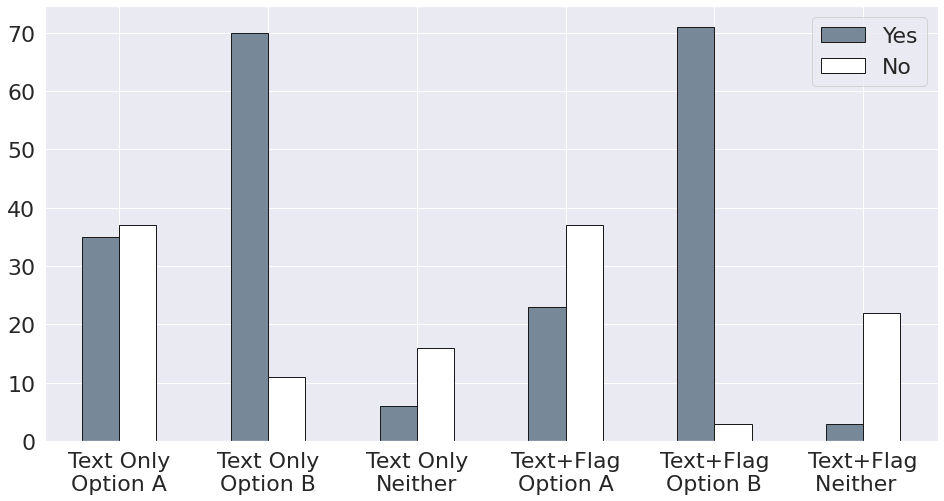}
  \caption{Dispells: \textbf{SPAM} Warning Tags}
  \label{fig:SPAM-Dispell}
\end{figure}

The A/B evaluation only obtained the preference for the \textbf{SPAM} warning tags without explicitly asking the participants to consider the security advice as applied to the tweets containing fabricated facts. To see if the \textbf{SPAM} warning tags actually work, we ask the participants to indicate if the tags helped them dispel fabricated facts in the example tweet. The results shown in Figure \ref{fig:SPAM-Dispell} indicate that the \textbf{SPAM} warning tags doesn't have to be users' best choice in order to work.

Roughly half of the ones that preferred the original warning tag commented that the text-only \textit{``helped them understand the meaning of the tweet in a broader context.''} In the text-and-flag variant participants found the warning tags helpful too rationalizing that \textit{``Twitter should just remove the whole post in general if it comes to a big red flag watermark.''} Even some of the neutral participants noted that the warning tag was reassuring  on the inaccuracy of the content. Overall, 62\% of the participants indicated that the \textbf{SPAM} warning tags worked for them with the desired effect of dispelling the fabricated effects of the COVID-19 vaccines. 

\subsubsection{Preferences and Political Leanings}
The COVID-19 pandemic didn't escape deep politicization and that naturally was reflected around the soft moderation effort of Twitter following the ban of President Donald Trump \cite{Pieroni}. We were interested, therefore, to see if participants' preferences are affected by their political leanings. For both \textbf{SPAM} warning tags variants, as indicated in Table \ref{tab:SPAM-politics}, the Pearson's Chi-Square test yielded a statistically significant relationship between their choices and where they stand on the political spectrum: $\chi(3) = 24.934$, $p = .000^{*}$ and $\chi(3) = 24.611$, $p = .000^{*}$, respectively. The original tags are appealing to left-learning with a 1:1 ratio to the moderate and 2:1 ratio to the right-leaning participants. The text-only \textbf{SPAM} variant has these ratios increased to a 4.5:1. Here, the neither ``A'' or ``B'' participants are uniformly distributed.

\begin{table}[htbp]
\renewcommand{\arraystretch}{1.0}
\small
\caption{\textbf{SPAM} vs Political Leanings}
\label{tab:SPAM-politics}
\centering
\begin{tabularx}{\linewidth}{|c|Y|Y|Y|Y|}
\Xhline{3\arrayrulewidth}
   & \multicolumn{4}{c|}{\textbf{Political Leanings}} \\\hline
\textbf{Option} & \textbf{Left} & \textbf{Moderate} & \textbf{Right} & \textbf{Apolitical}  \\\Xhline{3\arrayrulewidth}
 \multicolumn{5}{|c|}{\textbf{Text-only Warning Tags}} \\\Xhline{3\arrayrulewidth}
\textbf{Original (A)} & 27 & 27 & 14 & 0 \\\hline
\textbf{Spam (B)} & 54 & 12 & 13 & 0 \\\hline
\textbf{Neither} & 6 & 5 & 5 & 1 \\\Xhline{3\arrayrulewidth}
 \multicolumn{5}{|c|}{\textbf{Text-and-Flag Warning Tags}} \\\Xhline{3\arrayrulewidth}
\textbf{Original (A)} & 18 & 23 & 14 & 2 \\\hline
\textbf{Spam (B)} & 39 & 25 & 8 & 0 \\\hline
\textbf{Neither} & 2 & 4 & 7 & 2 \\\Xhline{3\arrayrulewidth}
 \end{tabularx}
\end{table}

The introduction of the flag tipped the left-leaning with a 1:1.3 ratio to the moderate and with a 1.3:1 ratio to the right-leaning ones that preferred the original warning tag. Left-leaning preferences for the text-and-flag \textbf{SPAM} warning tag were 1.56:1 with the moderates, but 4.875:1 with the right-leaning participants. The moderate and right-leaning were the most present for the neither ``A'' or ``B'' in the text-and-flag variant. Overall, the context is useful for the left-leaning and moderate participants the most, with a considerable portion of the moderates and right-leaning preferring a minimum intervention and distraction from Twitter.  

\subsubsection{Preferences and Demographic Identities}
The demographic identities, as the earlier evidence suggests \cite{cose2021, Zannettou}, factor in the way (mis)information is consumed from Twitter. Our analysis didn't find any significant relationship between the demographic identities and the preferences except between the education level and the text-and-flag \textbf{SPAM} variant: $\chi(3) = 17.328$, $p = .008^{*}$. The enhanced tag, as Table \ref{tab:SPAM-Education} reveals, roughly evenly splits the high school/GED and college graduates' preferences but almost entirely earns the preferences of the ones with a post-graduate degree. It also does so with a 3:1 ratio for the participants with some college degree. 

\begin{table}[htbp]
\renewcommand{\arraystretch}{1.0}
\small
\caption{\textbf{SPAM} vs Education Level}
\label{tab:SPAM-Education}
\centering
\begin{tabularx}{\linewidth}{|c|Y|Y|Y|Y|}
\Xhline{3\arrayrulewidth}
   & \multicolumn{4}{c|}{\textbf{Education Level}} \\\hline
\textbf{Option} & \textbf{High School/GED} & \textbf{Some College} & \textbf{College} & \textbf{Post-Graduate}  \\\Xhline{3\arrayrulewidth}
 \multicolumn{5}{|c|}{\textbf{Text-and-Flag Warning Tags}} \\\Xhline{3\arrayrulewidth}
\textbf{Original (A)} & 9 & 5 & 42 & 1 \\\hline
\textbf{Spam (B)} & 9 & 15 & 35 & 13 \\\hline
\textbf{Neither} & 2 & 1 & 11 & 0 \\\Xhline{3\arrayrulewidth}
 \end{tabularx}
\end{table}

\subsection{Improbable Interpretation of Facts (FFS)}
\subsubsection{A/B Evaluation}
The breakdown of preferences for both variants of the \textbf{FFS} warning tag variants are given in Table \ref{tab:FFS-Preferences}. In the text-only variant, only one third preferred the original tag and more than half of the participants choose the \textbf{FFS} text-only tag. Verbosity and confusion was the reason for almost two thirds of the participants to dislike the text-only \textbf{FFS} warning-tag. Roughly one third disliked it because of an anti-soft-moderation stance and one tenth provided no justification. 

The meaningful context provided by the \textbf{FFS} text-only warning tag (Figure \ref{fig:FFS-B}) was welcomed by almost 70\% of the participants \textit{``because it doesn't just say that the tweet is disputed, it mentions the various ways that the tweet is incorrect.''}. One out of ten participants liked that the \textbf{FFS} text-only warning tag because of the \textit{``assertive statement as opposed to just one word `disputed' in `A'. `B' is more specific.''}. A small number deemed the acronym as \textit{``funny/witty''} and 14.5\% simply just liked the \textbf{FFS} security advice.

\begin{table*}[!h]
\renewcommand{\arraystretch}{1.0}
\small
\caption{\textbf{FFS}: Preferences}
\label{tab:FFS-Preferences}
\centering
\begin{tabularx}{\linewidth}{|c|c|l|X|}
\Xhline{3\arrayrulewidth}
\textbf{Option} & \textbf{Pct.} & \textbf{Justifications} & \textbf{Representative Quotes}\\\Xhline{3\arrayrulewidth}
\multicolumn{4}{|c|}{\textbf{Text-only Warning Tags}} \\\Xhline{3\arrayrulewidth}

\multirow{3}{*}{\shortstack[c]{\textbf{Original (A)} \\ (33.5\%)}} & 60\% & Verbosity/confusion &  Because the other is just too many words. It just needs to be simple to understand. \\\cline{2-4}
& 31.6\% & Twitter intrusion & Twitter is bad enough when it tries to manipulate and control their own agendas. I don't want to see more additional information. \\\cline{2-4}
& 9.4\% & No justification & Warning Tag A. \\\Xhline{3\arrayrulewidth}

\multirow{4}{*}{\shortstack[c]{\textbf{FFS (B)} \\ (50.8\%)}} & 69.2\% & Meaningful Context & I would rather see context. It would bother me that some facts are missing and that I don't have the whole story. Vaccinations are too important of a topic to be misconstrued.  \\\cline{2-4}  
& 10\% & On-point warning &  Because it explains right off the bat that this content is manipulated or missing a source. \\\cline{2-4}

& 6.6\% & Attention Grabbing & ``B'' is engaging with the funny acronym.\\\cline{2-4}

& 14.2\% & No justification & Warning Tag B. \\\Xhline{3\arrayrulewidth}

\multicolumn{2}{|c|}{\textbf{Neither} (15.7\%)} & Neutral & I wouldn't utilize either.  \\\Xhline{3\arrayrulewidth}

 \multicolumn{4}{|c|}{\textbf{Text-and-Flag Warning Tags}} \\\Xhline{3\arrayrulewidth}
 \multirow{3}{*}{\shortstack[c]{\textbf{Original (A)} \\ (38.4\%)}} & 54.1\% & Distracting Flag & It doesn't have the big red watermark that might make people feel like victims. \\\cline{2-4}
 
 & 34.4\% & Twitter intrusion &  Human beings are simple creatures, and they do not respond well to being patronized. The latter is patronizing. \\\cline{2-4}
 
 & 11.5\% & No justification &  Warning Tag A \\\Xhline{3\arrayrulewidth}

\multirow{3}{*}{\shortstack[c]{\textbf{FFS (B)} \\ (46.0\%)}} & 47.9\% & Attention Grabbing & Option B really draws your attention and is impossible to miss or misunderstand. \\\cline{2-4}

 & 34.2\% & Meaningful Context &  It's important people really pick up on the fact this information might be misleading.  \\\cline{2-4}

 & 11\% & On-point warning &  The flag big and bold and it will tell me easily what to avoid and what is problematic. \\\cline{2-4}
 
 & 6.9\% & No justification &  Warning Tag B. \\\Xhline{3\arrayrulewidth}

\multicolumn{2}{|c|}{\textbf{Neither} (15.6\%)} & Neutral & I wouldn't prefer any of them. \\\Xhline{3\arrayrulewidth}
\end{tabularx}
\end{table*}

The preference against the text-and-flag \textbf{FFS} tag was again expressed in terms of destruction by more than a half of the participants preferring the original tag. A third of them cited the contempt for Twitter's decision to patronize users about how to interpret facts. A bit more than one tenth of the pro-original warning tags didn't provide justification. The participants pro the text-and-flag \textbf{FFS} liked the attention grabbing effect of the red flag noting that they \textit{``like that the red flag is big; You can see right away there is a problem with the tweet.''} in roughly half of the cases. The context (34.2\%) and the on-point warning that the tweet is a form of misinformation (11\%) was preferred because \textit{``knowing that something is missing context is more informative than knowing it's disputed; Everything is disputed by someone.''}. Only 6.9\% didn't provide justification pro the text-and-flag \textbf{FFS} warning tag. 

\subsubsection{Sentiment Analysis}
\begin{figure}[!h]
  \centering
  \includegraphics[width=\linewidth]{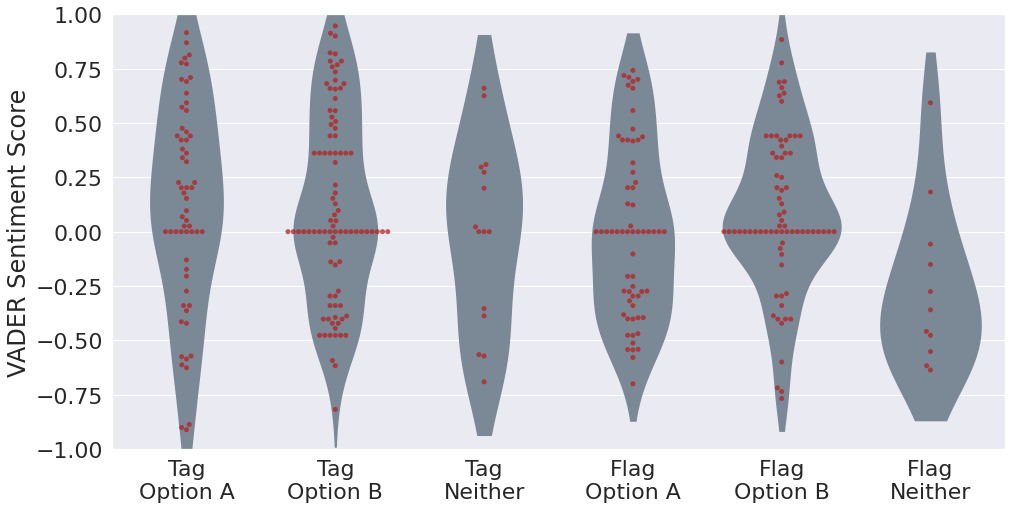}
  \caption{Sentiments: \textbf{FFS} Warning Tags. The violin plots show a multimodal distribution of sentiments (number of responses) over the VADER sentiment score ranging from -1 (negative) to +1 (positive) sentiment.}
  \label{fig:q7_swarm}
\end{figure}

The sentiment analysis of the preferences for \textbf{FFS} warning tags is shown in Figure \ref{fig:q7_swarm}. As the violin plots demonstrate, the original warning tag received roughly an equal number of positive sentiments for the \textit{``simple and straightforward and it doesn't try to make a judgment of the tweet''} as well as negative sentiments that \textit{``the red watermarking is overkill regardless of placement and size.''}. The justifications showed again showed low confidence (clout = 21.15) but positive emotions (tone = 67.72). The text-only \textbf{FFS} positive sentiment further outweighs the negative one praising the tag's way of \textit{``explaining why the facts are probably being used in a misleading way.''}. The praises show twice as more confidence as the ones for the original warning tag (clout = 51.93) and more positive emotions (tone = 72.32). 

The red flag in the \textbf{FFS} warning tag again caused a shift toward more negative sentiment as was cast as \textit{``condescending''} and \textit{``too distracting''}. The confidence plummeted in response to the flag-and-text variant
(clout = 19.3) with the emotions remaining negative (tone = 31.61). The positive sentiment is prevalent with the pro \textbf{FFS} text-and-flag tag participants, which wielded a tad better justifications (clout = 30.95) and expressed more positive emotions (tone = 60.52). The red flag again tilted the balanced sentiment of the neutral participants in the text-only variant toward a more negative one in the text-and-flag variant. 

\begin{figure}[!h]
  \centering
  \includegraphics[width=\linewidth]{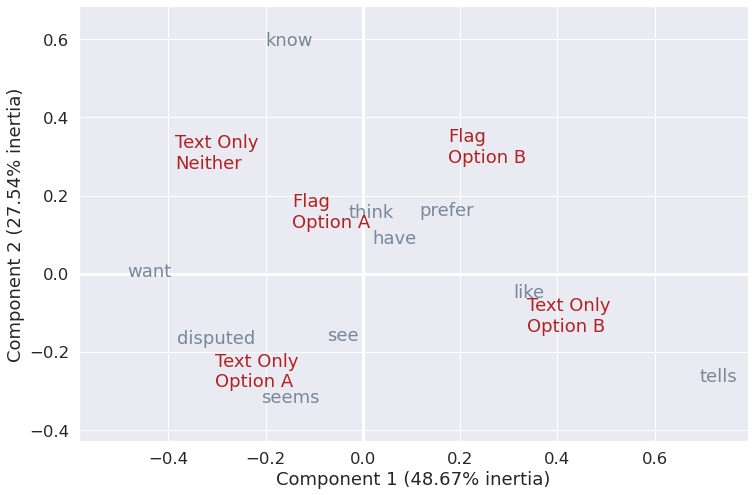}
  \caption{Correspondence Analysis: \textbf{FFS} Warning Tags}
  \label{fig:q7_corr}
\end{figure}

The CA for the \textbf{FFS} A/B evaluation is plotted in Figure \ref{fig:q7_corr}. For brevity, we used verbs as keywords here as the adjectives showed very similar dimensionality in the \textbf{SPAM} case (and vice versa). Here, the first component on the x-axis, accounting for 48.67\% of the variance in justifications, shows the responses in order of preference from left (least popular), to right (most popular). Verbs used in explanations for some of the less popular choices include ``disputed,'' and ``seems,'' which both are terms that indicate more ambiguity in the truth ( Option ``A'' in the text-only comparison), and ``know,'' which indicated more confidence ( Option ``A'' in the text-and-flag comparison). Justification for more popular responses include the verbs ``prefer,'' and ``like,'' which suggests approval for both \textbf{FFS} variants rather than a dislike for the original text-only warning. The most closely associated keyword with the text-only \textbf{FFS} warning tag is ``tells,'' which is an appreciation for the informal yet meaningful context conveyed. The y-axis, accounting for 27.54\% of the variance, shows the Option ``B'' preferences center around the origin as an indicator of higher consensus between the pro \textbf{FFS}.

\subsubsection{Dispelling Fabricated Facts}
\begin{figure}[htbp]
  \centering
  \includegraphics[width=\linewidth]{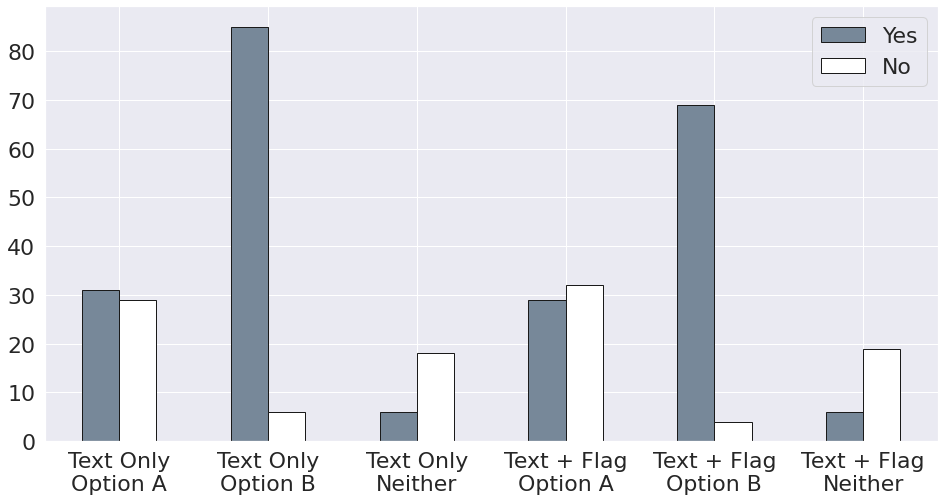}
  \caption{Dispels: \textbf{FFS} Warning Tags}
  \label{fig:FFS-Dispell}
\end{figure}

Figure \ref{fig:FFS-Dispell} shows that the detailed context provided through the \textbf{FFS} security advice is even more potent in dispelling improbable interpretation of facts. Overall, 68\% of the participants indicated that the \textbf{FFS} warning tags worked for them, which is a 6\% increase from the dispelling rate for the \textbf{SPAM} warning tags. Roughly half of the participants preferring the original tag conceded that the \textbf{FFS} warning tags in both variants are helpful in discrediting the manipulative tweet. A small but noticeable increase in the dispelling effect is also present for the neither ``A'' nor ``B'' participants compared to the \textbf{SPAM} warning tags. Similarly, the participants preferring both \textbf{FFS} tags were slightly more assertive of the desired effect compared to their responses for the \textbf{SPAM} tag.

\subsubsection{Preferences and Political Leanings}
The preferences for both \textbf{FFS} warning tags variants, as indicated in Table \ref{tab:SPAM-politics}, were related with a statistical significance to participants' political leanings: $\chi(3) = 27.732$, $p = .000^{*}$ and $\chi(3) = 36.483$, $p = .000^{*}$, respectively. The original tags are appealing to left-learning participants with a 1:1 ratio to the moderate ones and with a 2.5:1 ratio to the right-leaning participants. The text-only \textbf{FFS} variant has these ratios increased to a 3.6:1 between the left-leaning and the moderate participants and 4.8:1 between the left-learning and right-leaning participants. Unlike the \textbf{SPAM} variants, here, the neither ``A'' or ``B'' participants are dominantly right-leaning with a 2:1 ratio to both the left-leaning and moderate participants. 

\begin{table}[bhtp]
\renewcommand{\arraystretch}{1.0}
\small
\caption{\textbf{FFS} vs Political Leanings}
\label{tab:SPAM-politics}
\centering
\begin{tabularx}{\linewidth}{|c|Y|Y|Y|Y|}
\Xhline{3\arrayrulewidth}
   & \multicolumn{4}{c|}{\textbf{Political Leanings}} \\\hline
\textbf{Option} & \textbf{Left} & \textbf{Moderate} & \textbf{Right} & \textbf{Apolitical}  \\\Xhline{3\arrayrulewidth}
 \multicolumn{5}{|c|}{\textbf{Text-only Warning Tags}} \\\Xhline{3\arrayrulewidth}
\textbf{Original (A)} & 25 & 23 & 10 & 0 \\\hline
\textbf{Spam (B)} & 58 & 16 & 12 & 0 \\\hline
\textbf{Neither} & 4 & 5 & 10 & 1 \\\Xhline{3\arrayrulewidth}
 \multicolumn{5}{|c|}{\textbf{Text-and-Flag Warning Tags}} \\\Xhline{3\arrayrulewidth}
\textbf{Original (A)} & 20 & 26 & 12 & 2 \\\hline
\textbf{Spam (B)} & 38 & 23 & 7 & 0 \\\hline
\textbf{Neither} & 1 & 3 & 10 & 2 \\\Xhline{3\arrayrulewidth}
 \end{tabularx}
\end{table}

The introduction of the flag again kept the balance between the left-learning and moderate participants, but increased the ratio to almost 2:1 to the right-leaning ones that preferred the original warning tag. The left-leaning preferences for the text-and-flag \textbf{FFS} warning tag were in a 1.65:1 ratio with the moderates, but in an overwhelming 5.42:1 ratio with the right-leaning participants. The right-leaning again dominate in the neither ``A'' or ``B'' preferences for the \textbf{FFS} text-and-flag variant. Compared to the \textbf{SPAM} case, the extended \textbf{FFS} context is even more useful for the left-leaning participants. The moderates are roughly evenly split, but the right-leaning participants show a more salient anti-soft-moderation preference when exposed to the \textbf{FFS} warning tags.

\subsubsection{Preferences and Demographic Identities}
Same as before, only the level of education mattered when it comes to the preferences. The Pearson's Chi-square tests revealed a significant relationship in this case with $\chi(3) = 17.773$, $p = .007^{*}$. As the Table \ref{tab:FFS-Education} reveals, the high school/GED and the college graduates are slightly more in preference for the original tags, with a considerable dismissal for the soft-moderation altogether by the college graduates. The participants with some college-level education are 2:1 in ratio to the preference for the \textbf{FFS} text-and-flag variant with the ones preferring the original tag and 3:1 with the ones without a preference. The biggest difference is in for the participant with a post-graduate education level - they are almost entirely in favor of the \textbf{FFS} way of warning against improbable interpretations, manipulation, or selective choice of facts.    

\begin{table}[bhtp]
\renewcommand{\arraystretch}{1.0}
\small
\caption{\textbf{FFS} vs Education Level}
\label{tab:FFS-Education}
\centering
\begin{tabularx}{\linewidth}{|c|Y|Y|Y|Y|}
\Xhline{3\arrayrulewidth}
   & \multicolumn{4}{c|}{\textbf{Education Level}} \\\hline
\textbf{Option} & \textbf{High School} & \textbf{Some College} & \textbf{College} & \textbf{Post-Graduate}  \\\Xhline{3\arrayrulewidth}
 \multicolumn{5}{|c|}{\textbf{Text-and-Flag Warning Tags}} \\\Xhline{3\arrayrulewidth}
\textbf{Original (A)} & 11 & 6 & 42 & 1 \\\hline
\textbf{Spam (B)} & 8 & 12 & 35 & 13 \\\hline
\textbf{Neither} & 1 & 4 & 11 & 0 \\\Xhline{3\arrayrulewidth}
 \end{tabularx}
\end{table}


\section{Discussion}  \label{section:5}
In this study we were motivated to bring soft moderation closer to users' everyday experiences while minimizing imposition, which as witnessed, often backfire \cite{nspw2021}. We distinguished between a need for context when the hazard comes from the fabrication of facts and when the hazard comes from the interpretation of facts in a rather improbable way. In the first case, we were careful to avoid the perception trap of ``correction of feelings, not falshoods'' \cite{Lewandowsky} and used an analogy with spam emails. We did so because users, by now, can recognize spam when they see it \cite{Canfield} and accept that spam filtering, performed by email providers, works well \cite{Chachra}. Understanding this, we wanted to regain the trust in the provider - Twitter in the case of the warning tags - and signal absence of bias or judgment in their action \cite{Martel}.  

With this in mind, the \textbf{SPAM} warning tag shows a very promising step toward unified interpretation and increased trust in soft moderation (only related to COVID-19 misinformation, for now). If support from left-leaning participants was already hinted at from previous studies on soft moderation, it was nonetheless strongly reinforced in both the text-only (\textit{``...it tells participants, rather quickly, that the tweet is garbage''} and text-and-flag variants (\textit{``the red flag will alert me before I even read any of it''}. Moderates were evenly split, expectedly, but reassured that the text-only variant \textit{``really tells you more of what is going on''} while the text-and-flag variant \textit{``gives more specifics and is thus tougher to refute''}. In significant numbers, right-leaning participants made it clear that the text-only variant \textit{seems more appropriate because it's far more specific; the original tag feels more like an ad and nothing that I didn't already know.''} and praised the text-and-flag variant as \textit{``a large visual cue that's hard to ignore and will bring attention to the idea that something is going on with this information.''}. 

More promising evidence for the \textbf{SPAM} approach is the support across all levels of education without distinction of age, gender, or race/ethnicity. In the text-and-flag variant, only 10\% of the participant with only high school education/GED disliked both the ``A'' and ``B'' options while the rest gave equal support of 40\% for each option. Even though the participants with college diplomas tilted toward the option ``A'' (a relative difference of 8\%), the group unequivocally acknowledged that the text-and-flag \textit{'' is more clear and strong, and tells you exactly what is incorrect``}. After all, the \textbf{SPAM} tags helped more than 60\% of all the participants to dispel the fabricated facts about COVID-19 vaccine side effects.

In the second case, we wanted to avoid authoritative imposition and thus worded the warning not to personify senior public health experts, usually responsible for interpretation of facts \cite{Trethewey}. We also opted for a ``bold'' acronym choice to lure users' attention to the text of the warning tag, for a moment, instead of the warning tag as soft moderation. Once ``hooked,'' the cost to read the entire warning tag text was less then avoiding it as the derivation of  new meaning to acronyms and words is a pragmatic way of conveying context on social media - take for example the hashtags on Twitter \cite{Scott}. The text wasn't asking the user to ``get facts'' or ``learn more,'' but instead, it gave several convincing options for users themselves to pick why the context is fitting to the possibly misinformation tweet \cite{Redmiles16}.  

The \textbf{FFS} tag did just that and succeeded. Left-learning participants liked that the text-only variant \textit{``gives real reasons why this tweet is suspect''} and moderates seconded that the \textbf{FFS}'s context \textit{``goes more in depth and makes you more alert to the tweet''}. Right-leaning participants confirmed our idea to avoid any relationship to an imposing authority: \textit{``The context in `B' is better because Facebook came out saying that most if not all of their fact checkers don't check for facts, they just do it on opinion base. I'm sure Twitter does the same''}. The consensus across the participants of all political leanings that the \textit{``red flag watermark was really draws more attention''}, lead by the left-leaning ones, supports the potency of the \textbf{FFS} acronym as the ``hook'' entirely absent in the current soft moderation on Twitter. 

The \textbf{FFS} text-and-flag variant appealed almost entirely to all participants with a post-graduate level of education. Interestingly, they were concerned not just for themselves but other users on Twitter and misinformation in general, noting that \textit{``it's important people really pick up on the fact this information might be misleading''}. So were the participants with a college degree even though they again tilted toward the original tag ``A'': \textit{It seems like option `B' would help resolve the problems that false news or fake profiles create}. The participants with some college experience, in favor of the \textbf{FFS} tag, pointed out that a \textit{``disputed facts''} warning is less informative than a \textit{``missing facts''} warning. The support from the participants with only high school education/GED underlined the essential usability of the warning itself: \textit{``It makes it known that something is up with this post and I shouldn't trust it 100\% without doing more research.''}. Overall, the \textbf{FFS} achieved a 68\% effectiveness in dispelling an improbable interpretation of COVID-19 related facts among all participants.


\subsection{Observed Backfire Effect}
We did observe, albeit anecdotally, the backfire effect in 1.48\% of the participants' response (all politically right-leaning). One participant, who was pro enhanced warnings, even provided a testimony of the backfire effect: \textit{``I have seen some of my crazy friends of mine where they think if Twitter disputes it, then it makes it ever more correct.''}. The warning tags in the original option were blamed that \textit{``lead you to the lying, paid, `fact checkers'''}; The enhanced warning tags were dismissed because they \textit{``force an opinion on you and suppress a side that has been more accurate than the CDC and Fauci so far since COVID''}. Few participants even declared that the warning tags \textit{``makes them leave Twitter entirely''}, perhaps rappelled by the Twitter's  sweeping COVID-19 misleading policy from December 2021 \cite{TwitterPolicy}.

The contempt for Twitter's soft moderation was made clear in the responses of a considerable group of participants, stating that \textit{``Twitter is not a medial expert.''} Roughly a half preferring the original warning tags versus the \textbf{SPAM} variants cited Twitter intrusion into opinion formation as a choice for the ``lesser of the two evils.'' This fell down to a third of the Option ``A'' supporters in the case of the \textbf{FFS}, but considering that around 15\% of the overall participants did not have a preference for either of the options is an indication that soft moderation has still a lot to do to appear unbiased and non-judgmental to users with diverse perspectives \cite{Smith}.  

\subsection{User-tailored Soft Moderation}
Our results reveal several aspects worth considering for improving the soft moderation appeal among the Twitter users. There is no doubt that the meaningful context is useful but runs the risk of being avoided due to verbosity/confusion. In the \textbf{SPAM} case, a possible variation would be keeping just the acronym with a bit of text rewording, for example \textit{SPAM: Content like this usually ends in spam folders}. This improvement avoids the words ``strange,'' ``adverse'' and ``misinformation'' while indirectly hints that it should  be handled on user's discretion. Plus, it becomes a one-liner warning appearing more of a suggestion than an opinion voiced by Twitter, as several of our participants complained about. 

\begin{figure}[htbp]
  \centering
  \includegraphics[width=0.7\linewidth]{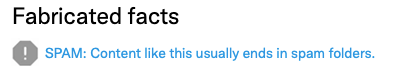}
  \caption{Update: Text-only \textbf{SPAM} Warning Tag}
  \label{fig:SPAM-Rework}
\end{figure}

Because the warning plays on the experience with spam emails, we also think it's worth testing the email iconography in line with the warning as shown in Figure \ref{fig:SPAM-Rework}. In this example, we borrowed the icon from Gmail's spam folder, but certainly could use any hexagram with an exclamation mark that provides contrast. This could also be an alternative to the red flag watermark to avoid participant recoiling from the sudden splash of red while still having an attention grabbing effect. Similarly, this could address the concerns for illegitimacy of the enhancement cited by some of the participants.  

\begin{figure}[htbp]
  \centering
  \includegraphics[width=\linewidth]{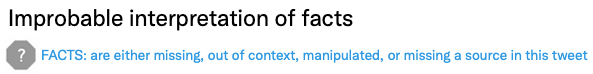}
  \caption{Update: Text-only \textbf{FACTS} Warning Tag}
  \label{fig:FFS-Rework}
\end{figure}

The context conveyed by the \textbf{FFS} tags was well received, but some participants, expectedly, expressed concerns about the ``catchy'' nature of the acronym. It is therefore worth testing dropping the acronym altogether or replacing it with simply the word ``facts'' as shown in Figure \ref{fig:FFS-Rework}. Here, the preceding iconography changes to a question mark, retaining the element of ``hook'' we envisioned in the first \textbf{FFS} variant. The following text is essentially the same, blending the acronym to look more ``professional,'' as the participants expected. 

The red flag watermark, the results confirm, produces the desired effect of attention-grabbing. However, adaptations could be made here too. Participants commented on the size and the transparency, so variations could include testing options where these two variables are determined by the level of confidence of fact checkers or the engagement it attracts over time, as suggested in \cite{nspw2021}. The watermark display could vary based on a particular user's content preferences, e.g. one group of users could see a red flag and another could see the words ``red flag'' as a watermark. Twitter already uses this approach to suggest adds and content in the users' feeds \cite{Bandy}.

\subsection{Ethical Considerations}
Ethical concerns do arise when dealing with misinformation, or allegedly harmful information, within the pluralistic social media population. The tension between impartiality, profitability, and social responsibility of the platforms might not always ensure that  misinformation is dealt with using consistent soft moderation criteria. With the honest, yet inevitable false-positives/negatives, the proposed enhancements - if applied - might be seen as unfair at best or simply harmful at worst. We therefore are open for democratic participation in the design that allows for remediation of concerns in such instances. Soft moderation, at least in our view, is a form of honest communicative action rather than an authoritative and absolute determination of truth, and as such, beneficial to all Twitter users without discrimination \cite{Andersen}. We are aware that facts change, become irrelevant, or are refuted over time so a retroactive application should also be considered to enable versatile soft moderation to the best of our (and Twitter's) abilities.

\subsection{Limitations}
We note several limitations of our study, which could be addressed in future work. The size of the sample could be enlarged to obtain an as varied as possible Twitter population. We used only two examples of misinformation on COVID-19, which is a limitation steaming both from restricted financial resources and limited attention span of participants \cite{Kaiser}. An extended, or perhaps a longitudinal study that incorporates more COVID-19 misinformation instances over a time could not just help generalize our findings, but reveal important behavioral patterns in dealing with soft moderation. Also, it could help with an A/B evaluation for warning tags pertaining other contested topics such as elections \cite{Zannettou}. 

Participants were exposed to generic formatting of the tweets emphasizing the content and the warning tags. In reality, misinformation could come from individual accounts, influencers, or accounts controlled by nefarious actors \cite{Sirivianos}. Misinformation is often amplified by social bots, and appears in users' feeds next to other posts, adds, with variable degree of visual interference \cite{Ferrara}. All of these aspects could influence the preferences for or against soft moderation. Controlling for them will require a study executed in partnership with Twitter where the enhancements are tested with selected users on the live platform.    

Such a test could not just capture the preferences of the regular Twitter users but help closely observe the ``backfiring,'' ``implied truth,'' and ``illusory truth'' effects. We didn't explicitly test for these in our study, but it is important to track how misinformation itself materializes in the individual Twitter consumption. Our A/B evaluation is limited by the current formatting of the original warning tags on Twitter \cite{Roth}. If Twitter chooses to reformat the tags, eliminate the links, or place them elsewhere, the enhancements also should change and the results might not hold for these new conditions.

\section{Conclusion}  \label{section:6}
This paper conveys the first extensive A/B evaluation of enhancements for misinformation warnings on Twitter. Providing users a meaningful context and attention-grabbing iconography, our results suggest, does help users recognize and contain COVID-19 misinformation. We weren't poised to solve the predicament of soft moderation in one shot; rather, the goal was to utilize the usable security body of knowledge to trace a path toward ``inoculation'' against information hazards on social media.

\bibliographystyle{ACM-Reference-Format}
\bibliography{\jobname}

\section*{Appendix}
\subsection*{Pre-Exposure Training}\label{sec:appendix-tweets}
\begin{figure}[!h]
  \centering
  \includegraphics[width=\linewidth]{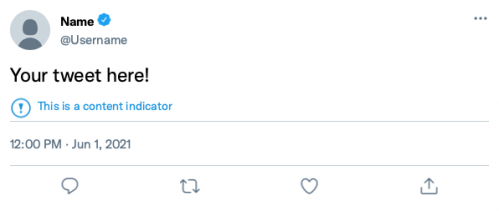}
  \caption{Generic Content Indicators on Twitter}
  \label{fig:content}
\end{figure}

Content indicator is defined as a label that is assigned by Twitter under a Tweet in blue font preceded by an exclamation mark as shown in the Figure \ref{fig:content}. Content indicators could be assigned for various types of contents, such as: misinformation, sensitive content, graphic content, etc.  



\end{document}